\documentclass[aps,twocolumn,nofootinbib,showpacs,final,balancelastpage,superscriptaddress]{revtex4}
\usepackage{graphicx}
\usepackage{tabularx}
\usepackage{amssymb}
\usepackage{amsmath}
\usepackage{amsbsy}
\usepackage{comment}
\usepackage{mathrsfs}
\usepackage{dsfont}
\usepackage{nicefrac}
\usepackage{wrapfig}
\usepackage{amsfonts}
\usepackage{url} 
\usepackage[nolist]{acronym}
\usepackage{nicefrac}
\usepackage{subfigure}
\usepackage{epsf,color,colordvi,pifont}
\usepackage{showkeys}

\usepackage{slashed}

\DeclareFontFamily{OT1}{pzc}{}
\DeclareFontShape{OT1}{pzc}{m}{it}{<-> s * [1.10] pzcmi7t}{}
\DeclareMathAlphabet{\mathpzc}{OT1}{pzc}{m}{it}

\begin{document}
\title{ Strong-field Breit-Wheeler pair production in $\mathrm{QED}_{2+1}$}
\author{A. \surname{Golub}}
\email{Alina.Golub@uni-duesseldorf.de }
\affiliation{Institut f\"{u}r Theoretische Physik I, Heinrich-Heine-Universit\"{a}t D\"{u}sseldorf, Universit\"{a}tsstra\ss{e} 1, 40225 D\"{u}sseldorf, Germany.}
\author{S. \surname{Villalba-Ch\'avez}}
\email{selym@tp1.uni-duesseldorf.de}
\affiliation{Institut f\"{u}r Theoretische Physik I, Heinrich-Heine-Universit\"{a}t D\"{u}sseldorf, Universit\"{a}tsstra\ss{e} 1, 40225 D\"{u}sseldorf, Germany.}
\author{C. \surname{M\"{u}ller}}
\email{c.mueller@tp1.uni-duesseldorf.de}
\affiliation{Institut f\"{u}r Theoretische Physik I, Heinrich-Heine-Universit\"{a}t D\"{u}sseldorf, Universit\"{a}tsstra\ss{e} 1, 40225 D\"{u}sseldorf, Germany.}

\begin{abstract}
The Breit-Wheeler pair production process in $2+1$ dimensional spacetime is investigated. In the perturbative regime, non-vanishing rates at the energy threshold are found when odd numbers of photons take part on the reaction. This behaviour is understood as a direct consequence of the reduced dimensionality and resembles a corresponding prediction made in gapped graphene monolayers. In the non-perturbative strong field regime the effect of the dimensionality manifests itself in a different rate dependence on the quantum non-linearity parameter. The consequence of this deviation is discussed briefly in line with the applicability of perturbation theory. We argue that, in addition to large values of the quantum non-linearity parameter, the superrenormalisable character of quantum electrodynamics in $2+1$ dimensional spacetime might give rise to a breakdown of perturbation theory within  certain energy scales. 
\end{abstract}


\date{\today}

\maketitle

\section{Introduction}
With the ongoing development of laser and accelerator infrastructures, the interest in producing electron-positron pairs from photon collisions has strongly grown over the last years as this process -- despite of being one of the most fundamental predictions of quantum electrodynamics ($\mathrm{QED}_{3+1}$) -- has not been validated experimentally in full glory. After the pioneering work due to Breit and Wheeler on the pair production by the collision of two photons \cite{BreitWheeler} (linear Breit-Wheeler process), further theoretical efforts revealed a phenomenology which manifests --among other issues-- the non-linear feature of $\mathrm{QED}_{3+1}$. Initially, several authors considered interaction of light quanta with a plane-wave electromagnetic background \cite{ReissBW, NikishovRitus1, NikishovRitus2, RitusReview}. Special attention has been paid to the   perturbative [$\xi \ll 1$] and non-perturbative [$\xi \gg 1$] regimes defined by the intensity parameter of the involved laser wave $\xi = -\mathpzc{e}E_0/(\omega m)$  with electron charge $\mathpzc{e}<0$, its  mass $m$, frequency $\omega$ and amplitude $E_0$ of the plane wave.\footnote{From now on we use a metric with signature $\mathrm{diag}(g^{\mu \nu})= (1,-1,-1)$. Throughout, a natural unit system where $c=\hbar=1$ is adopted.} Besides, more elaborated treatments have been put forward in order to reach a more precise description of laser configurations \cite{EhlotzkyReview, diPiazzaReview, NarozhnyReview, Heinzl, Krajewska2012, Kaempfer2012, Kaempfer2016,  Jansen2016, diPiazza, Grobe, Hartin} with particular emphasis on the highly non-linear non-perturbative regime due to important links to a wider range of elusive phenomena 
\cite{Krajewska2014, Selym, Jansen2013,  Meuren, Blackbourn,Kaempfer2020}. 
Of special interest has been the region in which the quantum non-linearity parameter  $\chi = \xi kk'/m^2$ --involving photon momenta $k$ and $k'$-- exceeds unity considerably as it might come accompanied with a breakdown of perturbation theory when $\alpha \chi^{2/3} \geq 1$ ($\alpha=\mathpzc{e}^2/4\pi \approx 1/137$ is the fine structure constant)  \cite{Baumann2, diPiazza2, Blackburn2, Yakimenko,NikishovRitus3,Mironov, Podszus, Ilderton}.

Despite the theoretical endeavour, experimental validation of Breit-Wheeler pair creation has so far been possible solely in the multiple-photon regime at $\xi \lesssim 1$ \cite{SLAC}. While this experiment provided important insights into this process, neither linear pair creation --for recent proposals see \cite{KingGies, Pike, Drebot, Ribeyre,Wang, Golub}-- nor Breit-Wheeler pair creation at $\xi \gg1$ have been observed by now. The reason why the former process is still elusive stems from the current limitations of reaching energetic photon beams with adequate intensities. Conversely, accessing to the non-perturbative regime demands peak intensities $I = E_0^2 \sim I_c \approx 4.6 \times 10^{29} \ \mathrm{Wcm^{-2}}$, corresponding to the Schwinger electric field scale $E_c\approx 1.3 \times 10^{16} \ \mathrm{Vcm^{-1}} $ in the relevant frame of reference. Reaching this condition requires ultrahigh laser intensities $ I \sim (10^{21}-10^{24}) \ \mathrm{Wcm^{-2}}$ in the laboratory frame in conjunction with a beam of sufficiently energetic $\gamma$ photons, which represents a hard challenge to overcome (for recent proposals see \cite{LUXE, FACET}).

A low-energy test ground for yet unobserved $\mathrm{QED}_{3+1}$ processes is provided by bandgapped graphene monolayers. In this two dimensional honeycomb of carbon atoms charge carriers near the degeneracy points possess a Dirac-like dispersion relation with the speed of light replaced by the Fermi velocity $v_F \approx 1/300$. Hence, their behaviour can be effectively described by a 2+1 dimensional Dirac model and the interband transition of electrons, i.e. the production of electron-hole pairs, can be exploited as a toy model for addressing $\mathrm{QED}_{3+1}$ pair production-related questions. This idea has motivated investigations of analogues to the Schwinger mechanism \cite{SelymGraphene, Graphene} by delivering inherent insights caused by the low dimensional material: the power of the pre-exponential factor is changed from $2$ in $\mathrm{QED}_{3+1}$ to $3/2$ in graphene. Likewise, bandgapped graphene has been presented as a suitable material to simulate the Breit-Wheeler process with two (linear case) and three photons \cite{GolubGraphene}. In the latter case, as in the studies linked to the Schwinger-like mechanism, inherent properties due to the low dimensional character of the medium came to a scene. Contrary to $\mathrm{QED}_{3+1}$, the rate linked to this process does not vanish at the energy threshold. This poses special feature questions  whether its occurrence extends to perturbative Breit-Wheeler reactions in which more than  three photons are absorbed  and  to which extent  the Lorentz symmetry breaking caused by  the Fermi velocity takes part in its realization. Addressing these questions is not an easy task within the Dirac model,  mainly because   the dispersive properties of the medium  make the theoretical treatment difficult  to  handle.\footnote{As a consequence of the asymmetry introduced by the simultaneous appearance of the Fermi velocity and the speed of light, solutions to the Dirac equation of graphene in the field of a plane wave are much more difficult to obtain than in $\mathrm{QED}_{3+1}$.}

Against this background, we study in the present paper the Breit-Wheeler pair production in 2+1 dimensional quantum electrodynamics ($\mathrm{QED}_{2+1}$) in various interaction regimes from weak to very strong fields. In particular we show that the threshold lifting persists in this Lorentz invariant framework when an odd number of photons are absorbed, confirming that this phenomenon results as an inherent consequence of the reduced spacetime dimensionality. Besides, we reveal characteristic influences of the dimensionality on the non-perturbative production rates for both small and large values of the quantum non-linearity parameter, and discuss the implications for a breakdown of perturbation theory in this scenario.

Considerations of low-dimensional scenarios have proven useful in various areas of physics. They revealed, for example, valuable insights on non-perturbative aspects of QCD (see e.g. \cite{Fister} and references therein) and even provided a solvable quantum field theory model for gravity \cite{Banks}. Investigations of the Chern-Simons term --intrinsically linked to $\mathrm{QED}_{2+1}$ \cite{Templeton}-- have improved our understanding of the high temperature behaviour of $\mathrm{QED}_{3+1}$, leading in addition to an accurate description of the quantum Hall effect \cite{Prange}. Likewise, various research on the subject of Schwinger pair creation in arbitrary spacetime dimensions have been performed by considering different field configurations such as constant uniform electric and magnetic fields \cite{Lin} and electric fields of finite duration \cite{GavrilovGitman}. Regarding Breit-Wheeler pair production, a recent study based on a model Hamiltonian in one spatial dimension has allowed for spacetime resolution of this process \cite{LuBW}.

Another important area of applicability for lower dimensional theories is the field of quantum simulation. Gradually, this branch is gaining interest as it enables to simulate the behaviour of various many-body ensembles via ultracold atom systems placed in sophisticated optical lattices. Because of a complex technical implementation, these quantum simulators have been restricted so far to systems of lower dimensionality, so that the question arises, which differences occur as compared to 3+1 dimensions. In the realm of quantum electrodynamics, the progress in this field has led to an experimental implementation of the 1+1-dimensional Schwinger mechanism \cite{Martinez,Pinero}, which was investigated thoroughly by theoreticians \cite{Schutzhold1,Schutzhold2,Pichler}, whereas current research efforts are spent towards $\mathrm{QED}_{2+1}$ \cite{Zohar,Zache,Schutzhold3,Ott}. Hence, with the present study of the non-perturbative Breit-Wheeler pair creation we provide a scenario which may be further explored through the branch of quantum simulation.

Our paper is organised as follows: in Sec.~\ref{sec:Volkov} we introduce Volkov states in $\mathrm{QED}_{2+1}$ and apply them in Sec. \ref{BW}  to obtain a general expression of the Breit-Wheeler pair production rate. Afterwards, in Sec. \ref{Asymptotics} we evaluate the rate numerically in various intensity regimes and support this analysis by an asymptotic study. Our conclusions are given in Sec.\,V, while  the two-photon pair production process is described in the lowest order of perturbation theory in Appendix \ref{pairCrePert}.

\section{Volkov states in 2+1 dimensions}\label{sec:Volkov}
In a 2+1 dimensional spacetime the time evolution of planar relativistic electrons interacting with an electromagnetic field $a_\mu(x)$ is described by the Dirac equation
 \begin{equation}\label{diracEq}
(i\slashed{D} -m)\psi=0,
\end{equation}
which manifests a $\mathrm{SO}(1,2)$ invariance \cite{GuilarteMayado, Bellucci}. In this expression $\psi(x)$ stands for a two component spinor wave function, whereas $\slashed{D} \equiv \gamma^\mu D_\mu$ and  $ D_\mu=\partial_\mu +iea_\mu$ refers to the covariant derivative with $e<0$ denoting the electric charge in $2+1$ dimensions. Observe that the latter  notation differs from  the one used in the introduction ($\mathpzc{e}$). The reason behind this change will be discussed below. The $\gamma^\mu$-matrices linked to this low dimensional scenario are determined by Pauli matrices $\gamma^\mu = (\sigma_3, i\sigma_2,-i\sigma_1)$ and are also constricted by the Clifford-algebra $\{ \gamma^\mu, \gamma^\nu\}=2g^{\mu \nu}\mathds{1}_{2\times 2}$ with $\mu=0,1,2$.

As in 3+1 dimensions the solvability of Eq.~\eqref{diracEq} is restricted to a certain class of electromagnetic fields. Here, we will focus on the solution resulting when $a^\mu(x)$ is chosen as a polarized plane-wave-like three-potential
\begin{equation}\label{elmfield}
a^\mu(kx)=a_0\epsilon^\mu \mathrm{cos}(kx)
\end{equation}  
characterised by the amplitude $a_0$, the wave vector $k^\mu=(k_0,\vec{k})$ and a transverse polarisation $\epsilon^\mu$ [$\epsilon^\mu k_\mu =0$], which is normalised according to $\epsilon^\mu\epsilon_\mu=-1$. We remark that in 2+1 dimensions the magnetic field provided by $a^\mu(x)$ is a pseudo scalar, whereas the electric field is a two component vector. It is worth emphasising that the field in Eq.~\eqref{elmfield} is supposed to be a solution of a Maxwell equation without taking the Chern-Simons contribution into account, an assumption which applies  whenever the energy-momentum transfer is much larger than the topological mass of the gauge field.

Despite of the inherent differences caused by the dimensionality the procedure for establishing the 2+1 dimensional Volkov-states does not differ much from the well known  approach in $\mathrm{QED}_{3+1}$ (see for instance Refs. \cite{Volkov,LandauLifshitz,NikishovRitus1}). In line, we find 
\begin{multline}\label{Volkov}
\psi_{p^-}(x)=\sqrt{\frac{m}{q_0^{-}A}}\left(1+\frac{e}{2kq^{-}}\slashed{k}\slashed{a}\right)u_{p^{-}}\times \\
\mathrm{exp}\left[-iq^{-}x-i\frac{ea_0q^{-}\epsilon}{kq^{-}}\mathrm{sin}(kx) -i\frac{e^2a_0^2}{8kq^{-}}\mathrm{sin}(2kx)\right].
\end{multline}
The state given above has been normalised in such a way that the time averaged electron density $\langle j^0 \rangle = \langle \bar{\psi}_{p^-}\gamma^0\psi_{p^-}\rangle$ with $\bar {\psi }_{p^-}\equiv \psi _{p^-} ^{\dagger }\gamma ^{0} $ amounts to one particle in the normalisation area $A$. While $u_{p^{-}}$ refers to a free spinor [$a_\mu=0$] with positive energy, the quantity $q^{-}_\mu = p^{-}_\mu +\frac{e^2a_0^2}{4kp^{-}}k_\mu$ stands for an averaged effective electron momentum with $kp^-=kq^-$, $\epsilon p^-=\epsilon q^-$, where the electron momentum is denoted by $p^-$. Moreover, the Volkov solution for a positron $\psi_{p^+}(x)$ can be read off from Eq.~\eqref{Volkov} by carrying out the replacements: $p^-\to-p^+$ and $u_{p^{-}} \to v_{p^{+}} $, where $v_{p^{+}}$ is the free negative-energy solution. Both spinors $u_{p^{-}}$ and $v_{p^{+}}$ are normalised in accordance with the following rules
\begin{equation}\label{spinorProperties}
u_{p^{-}}\bar{u}_{p^{-}}=\frac{\slashed{p}^- +m}{2m}, \quad v_{p^{+}}\bar{v}_{p^{+}}=\frac{\slashed{p}^+  -m}{2m} 
\end{equation}
and $\bar{u}_{p^{-}}\gamma_\mu u_{p^{-}}=p^-_\mu/m$ for $\bar{u}_{p^{-}} \equiv u^\dagger_{p^{-}}\gamma^0$, $\bar{v}_{p^{+}} \equiv v^\dagger_{p^{+}} \gamma^0$.

\section{Breit-Wheeler pair creation in $\mathrm{QED}_{2+1}$}\label{BW}
\subsection{General considerations}
\begin{figure}
\includegraphics[width=5cm]{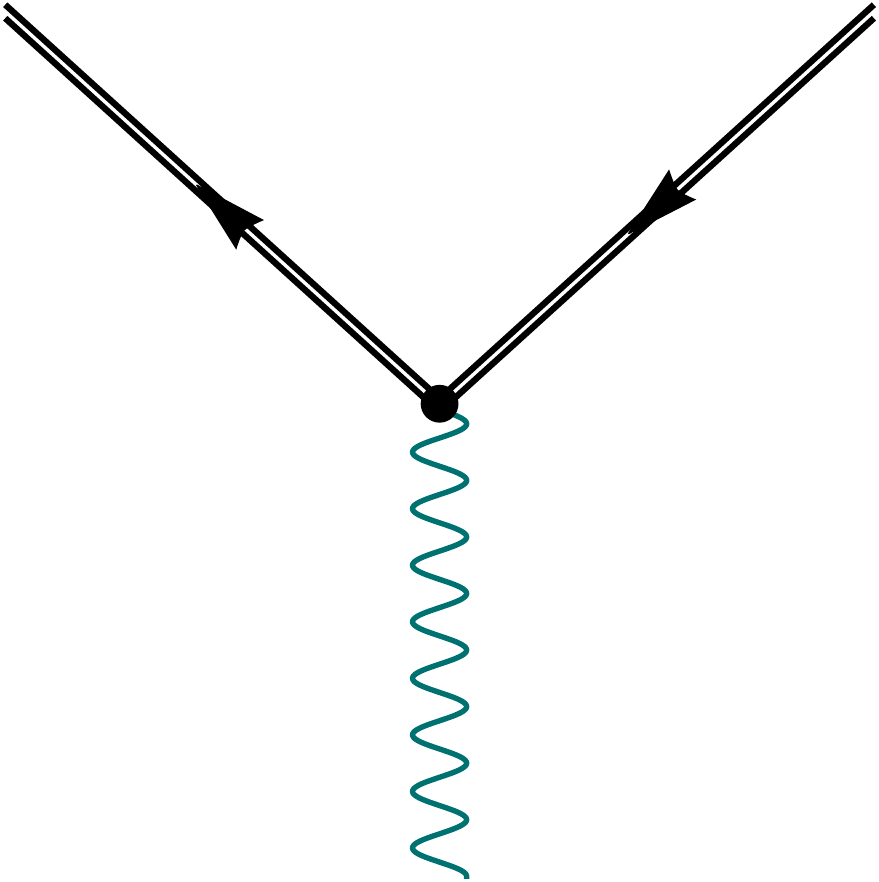}
\caption{\label{FeynBW} Feynman diagram of the Breit-Wheeler pair creation process in a plane-wave background. Here, the wavy line represents a quantised photon, whereas the solid double lines stand for Volkov states of electron and positron.}
\end{figure}

In $\mathrm{QED}_{2+1}$, the reduction of space dimensions gives rise to changes in the physical dimensions of the involved fields $\psi$, $a_\mu$ and the electric charge $e$
\begin{equation}
[\psi] = \mathcal{E}, \ [a] = \mathcal{E}^{1/2} \ \mathrm{ and}  \ [e] = \mathcal{E}^{1/2},
\end{equation}
where $\mathcal{E}$ stands for the dimension of energy. 
As the charge in the equation above has a positive energy dimension, $\mathrm{QED}_{2+1}$ belongs to the class of super-renormalisable theories. This fact prevents us from identifying $e^2/4\pi$ with the traditional fine-structure constant $\alpha \approx 1/137$ and demands to look for an adequate dimensionless parameter from which a perturbative treatment can be justified.

In the scope of the strong-field Breit-Wheeler process two gauge fields are involved in the pair creation: a strong field $a_\mu$ and a low-intensity field $a'_\mu$. Hence, we can identify dimensionless\footnote{It is worth remarking that, in a minimally coupled framework, the combination $e a^{(\prime)}_0$ has always a dimension of energy regardless the number of space dimensions $d$: $[e a^{(\prime)}_0] = \mathcal{E}^{(3-d)/2}\mathcal{E}^{(d-1)/2}= \mathcal{E}$.} intensity parameters $\eta =-e a_0/m>0$ and $\eta ^{\prime}=-e a^{\prime}_0/m>0$, where $a_{0}$ and $a^{\prime}_{0}$ stand for the field amplitudes, and perform the perturbative expansion in $\eta'$ assuming that $\eta' \ll 1$. 

Accordingly, our starting point is the  generic transition amplitude for the Breit-Wheeler process (see Fig.~\eqref{FeynBW}): 
\begin{equation}\label{transAmpl}
S_{fi}=-ie\int d^3x\bar{\psi}_{p^-}(x)\slashed{a}'(x)\psi_{p^+}(x),
\end{equation}
where $a'_\mu(x)=\frac{a'_0}{2}\epsilon'_\mu\mathrm{e}^{-ik'x}$ denotes the photon wave function, which refers to the amplitude $a_0'$ rather than being normalised to one particle in the area, and the polarisation $\epsilon'_\mu$ satisfies the conditions $\epsilon' k' =0, \epsilon' \epsilon'=-1$. In this expression $\psi_{p^-}(x)$ and $\psi_{p^+}(x)$ stand for the Volkov solutions as given in Sec.~\ref{sec:Volkov}. 

However, we point out that the applicability of Eq.~\eqref{transAmpl} also depends on the regime dictated by $\eta$, where additional specification of the expansion parameters may be in order. Indeed, as it happens in $\mathrm{QED}_{3+1}$, large values of $\eta \gg 1$  might compensate  the smallness of $\eta'$, making the effective coupling strength far from being  perturbative.  In this scenario the effect of radiative corrections should be included in the transition amplitude. Further discussion on this topic will follow at the end of Sec.~\ref{EtaLarge}.

To evaluate the integral in Eq.~\eqref{transAmpl}, we follow the usual procedure and introduce a Fourier expansion of an exponential
\begin{equation}
\mathrm{e}^{-iz_-\mathrm{sin}(kx)-iz_+\mathrm{sin}(2kx)}=\sum_{n=-\infty}^\infty \mathrm{e}^{-inkx} \tilde{J}_n(z_-,z_+),
\end{equation}
in which the generalised Bessel function is defined as an infinite sum of a product of ordinary Bessel functions of one argument \cite{Krainov, Reiss, NikishovRitus1}
\begin{equation}\label{Bessel}
\tilde{J}_n(z_-,z_+)=\sum_{m=-\infty}^\infty J_{n-2m}(z_-)J_m(z_+).
\end{equation}  
and write ($l=1,2$)
\begin{equation}
\begin{split}
&\mathrm{cos}^l(kx)\mathrm{e}^{-iz_-\mathrm{sin}(kx)-iz_+\mathrm{sin}(2kx)}=\sum_{n=-\infty}^\infty \mathrm{e}^{-inkx} \tilde{\mathcal{J}}^l_n, \\
&\tilde{\mathcal{J}}^1_n \equiv \frac{1}{2}\left[ \tilde{J}_{n+1}(z_-,z_+) + \tilde{J}_{n-1}(z_-,z_+)\right], \\
&\tilde{\mathcal{J}}^2_n \equiv \frac{1}{4}\left[\tilde{J}_{n+2}(z_-,z_+) + 2\tilde{J}_n(z_-,z_+)+\tilde{J}_{n-2}(z_-,z_+)\right].
\end{split}
\end{equation}
By exploiting these definitions we can write Eq.~\eqref{transAmpl} as 
\begin{multline}\label{matrix1}
S_{fi}=-i\frac{ea'_0}{2}\sqrt{\frac{m^2}{q_0^+q_0^-A^2}}\sum_{n=-\infty}^\infty (2\pi)^3\delta_{nk+k',q^++q^-}\\
\times \bar{u}_{p^{-}}M_n v_{p^{+}}, 
\end{multline}
where $\delta_{x,y}\equiv \delta^3(x-y)$ and $M_n$ is of the following form
\begin{multline}
M_n=\slashed{\epsilon}'\tilde{J}_n + \frac{ea_0}{2}\tilde{\mathcal{J}}^1_n \Big({\frac{\slashed{\epsilon}\slashed{k} 
\slashed{\epsilon}'}{kq^{-}} -\frac{\slashed{\epsilon}'\slashed{k} 
\slashed{\epsilon}}{kq^{+}}}\Big)\\
-\frac{e^2a_0^2}{2kq^{+}kq^{-}}\epsilon'^\mu k_\mu \tilde{\mathcal{J}}^2_n \slashed{k}.    
\end{multline}
Notice that energy-momentum conservation $q^++q^-=nk+k'$, anticommutativity of gamma matrices as well as transversal condition have been employed and the arguments of generalised Bessel functions read $z_-=e a_0(\frac{q^+\epsilon}{q^+k}-\frac{q^-\epsilon}{q^-k})$, $z_+=-\frac{e^2a_0^2}{8}\frac{kk'}{kq^+kq^-}$.
Next, we consider the differential rate of pair creation per area defined as
\begin{equation}\label{rateinitial}
dR=\frac{|S_{fi}|^2}{TA} A^2\frac{d^2q^-}{(2\pi)^2}\frac{d^2q^+}{(2\pi)^2}.
\end{equation}
Here, the phase space is expressed in terms of the effective particle momenta, in accordance with the normalization chosen in Eq.~\eqref{Volkov}. When inserting Eq.~\eqref{matrix1} into the equation above we end up with 
\begin{multline}\label{rateM}
dR=\frac{e^2a_0'^2m^2}{8\pi}\sum_{n=-\infty}^\infty |\bar{u}_{p^{-}}M_n v_{p^{+}}|^2\\
\times \delta_{nk+k',q^++q^-} 
 \frac{d^2q^-}{q_0^-} \frac{d^2q^+}{q_0^+},
\end{multline}
where the sum over $n$ can be interpreted as a sum over a number of absorbed photons from the classical field $a^\mu(kx)$. Furthermore, the
square of the amplitude $|\mathcal{M}^n|^2=|\bar{u}_{p^{-}}M_n v_{p^{+}}|^2$ can be written as
\begin{equation}\label{matrixsquared}
|\mathcal{M}^n|^2 = \mathrm{Tr}\left[\frac{\slashed{p}^- +m}{2m}M_n \frac{\slashed{p}^+  -m}{2m} \bar{M}_n\right]
\end{equation}
with $\bar{M}_n = \gamma^0M^{\dagger}_n\gamma^0$. To show the relation above the properties of free Dirac spinors in 2+1 dimensions given in Eq.~\eqref{spinorProperties} were used. It is interesting to point out that, although the form of the squared amplitude remains unchanged as compared to 3+1 dimensions, the trace is no longer taken over a $4\times4$ matrix, but over one of the dimension $2\times2$. This feature is responsible for introducing major differences with respect to $\mathrm{QED}_{3+1}$.\footnote{ In the calculations we have used \label{footnote4}
\begin{align*}
&\gamma^0\gamma^{\dagger\mu}\gamma^0=\gamma^\mu,\quad \gamma^\mu\gamma_\mu = 3\mathds{1}_{2\times 2},\quad \mathrm{Tr}[\gamma^\mu\gamma^\nu]=2g^{\mu \nu}, \\
&\mathrm{Tr}[\gamma^\mu\gamma_\mu]=6, \quad \gamma^\mu\gamma^\nu \gamma_\mu = -\gamma^\nu, \quad \mathrm{Tr}[\gamma^\nu]=0, \\
& \mathrm{Tr}[\gamma^\mu\gamma^\nu \gamma^\alpha]=-2i\epsilon^{\mu \nu \alpha}, \quad \gamma^\mu \gamma^\nu \gamma^\alpha \gamma_\mu = 4g^{\nu \alpha}-\gamma^\nu \gamma^\alpha,\\
& \gamma^\mu \gamma^\nu \gamma^\alpha \gamma^\beta \gamma_\mu = \gamma^\nu \gamma^\alpha \gamma^\beta -2\gamma^\beta \gamma^\alpha \gamma^\nu ,\\
&\mathrm{Tr}[\gamma^\mu\gamma^\nu \gamma^\alpha\gamma^\beta]=2(g^{\mu\nu}g^{\alpha\beta}-g^{\mu\alpha}g^{\nu\beta}+g^{\mu\beta}g^{\nu\alpha})
\end{align*}
with Levi-Civita tensor $\epsilon^{\mu \nu \alpha}$ with $\epsilon^{012}=1$. }

Additionally, since the gauge field in 2+1 dimensions has only one transverse polarization direction, we can make use of the relation 
\begin{equation}\label{gmunu}
\epsilon'^\mu \epsilon'^\nu = -g^{\mu \nu}-\frac{k'^\mu k'^\nu -n'k'(k'^\mu n'^\nu + k'^\nu n'^\mu )}{(n'k')^2},
\end{equation} 
where $n'^\mu =(1,0,0)$ and $k'^\mu=(\omega', \vec{k}')$. It is worth emphasising that, owing to the Ward identity, only the term containing $-g^{\mu \nu}$ will contribute to the squared amplitude. Having these properties in mind we arrive at 
\begin{equation}\label{matrixfinal}
|\mathcal{M}^n|^2 =\tilde{J}_n^2 -\frac{e^2a_0^2}{m^2}\left(1-\frac{(kk')^2}{4kq^+kq^-}\right)\left[ (\tilde{\mathcal{J}}^1_n)^2 - \tilde{J}_n\tilde{\mathcal{J}}^2_n\right],
\end{equation}  
where the energy-momentum balance and transversal condition ($k\epsilon=0$) have been used. We remark that, when deriving this formula the following relation was exploited \cite{Reiss, NikishovRitus1}
\begin{equation}
(n+2z_+)\tilde{J}_n -z_-\tilde{\mathcal{J}}^1_n -4z_+\tilde{\mathcal{J}}^2_n =0.
\end{equation}
Inserting Eq.~\eqref{matrixfinal} into Eq.~\eqref{rateM} the rate for Breit-Wheeler pair creation in 2+1 dimensions reads
\begin{multline}\label{rate}
dR=\frac{e^2a_0'^2m^2}{8\pi}\sum_{n \geq n_0}^\infty \delta_{nk+k',q^++q^-} \frac{d^2q^-}{q_0^-} \frac{d^2q^+}{q_0^+}\\
\times \left[ \tilde{J}_n^2 -\eta^2\left(1-\frac{(kk')^2}{4kq^+kq^-}\right)\left[ (\tilde{\mathcal{J}}^1_n)^2 - \tilde{J}_n\tilde{\mathcal{J}}^2_n\right]\right].
\end{multline} The fact that the summation index starts at $n_0=2m_{*}^2/(kk')$ with $m_*^2=m^2(1+\eta^2/2)=q^{\pm 2}$ referring to the effective electron mass is understood here as a consequence of the energy-momentum balance. Leaving aside the lower dimensionality of the delta functions and the involved integration measures, the main source of difference between $dR$ and the unpolarized rate in 3+1 dimensions (see for instance Eq.~(15) in \cite{NikishovRitus1}) lies in the precise structure of the squared amplitude (compare with Eq.~\eqref{matrixfinal}). It resembles the squared amplitude of the pair production by a photon, polarization of which is parallel to the electric field of the wave (as given in Eq.~(35) in \cite{NikishovRitus1} or in Eq.~(3) in \cite{NikishovRitus2}). This fact highlights the restriction provided by a lower dimensionality: both photon wave vectors and polarisations are bound to a plane. Hence, when moving the system to a center-of-momentum frame, where photons are counterpropagating, their polarisations have to be parallel.

 In order to highlight the consequences linked to the spacetime dimensionality further, we change the integration variables to a set referred to the center-of-momentum of the created particles, where $\vec{q}^{\ -}=-\vec{q}^{\ +}=-\vec{q}$, $n\vec{k}=-\vec{k}'$, $n\omega=\omega'$. Afterwards, $q^-$ is integrated out and the remaining integral is expressed in polar coordinates 
\begin{multline}\label{ratecmf}
R=\frac{e^2a_0'^2m^2}{8\pi}\sum_{n\geq n_0}^\infty \int_{0}^{2\pi}d\phi \int_0^{\infty} \frac{dq}{2q_0}\delta(q-q_n^*)\\
\times\left( \tilde{J}_n^2 -\eta^2\left(1-\frac{1}{1-\frac{q^2}{\omega'^2} \mathrm{cos}^2(\phi)}\right)\left[ (\tilde{\mathcal{J}}^1_n)^2 - \tilde{J}_n\tilde{\mathcal{J}}^2_n\right]\right)
\end{multline}
when setting $|\vec{q} \ | \equiv q$, $kq^+=\omega[\omega'-q\mathrm{cos}(\phi)]$, $kq^-=\omega[\omega'+q\mathrm{cos}(\phi)]$, $\epsilon q^- = -\epsilon q^+ = q \mathrm{sin}(\phi)$ and $q_n^*=\sqrt{\omega'^2-m_*^2}$, $q_0=\omega'$. Additionally, we introduce a Mandelstam invariant $s=\sqrt{kk'/2m^2}$ which allows us to write $q_n^*=\omega'\sqrt{1-(1+\eta^2/2)/ns^2}$. Requiring that the number of absorbed photons exceeds a critical number $n\geq (1+\eta^2/2)/s^2$, we obtain for every  value of $n$ a threshold energy $s^2 \geq (1+\eta^2/2)/n = s_n^2$ which must be overpassed in order that the process takes place.

\subsection{Threshold behaviour}
Now, we are interested in investigating the behaviour of the rate at the energy threshold, where the particles  are created with zero momentum ($q=0$). It is noteworthy that at the threshold the integral over the phase space will not always vanish and is governed solely by the form of the amplitude. This fact represents a crucial difference to the 3+1 dimensional case where after integrating over $q^-$ in the center-of-momentum frame the remaining integral $\int \frac{d^3q}{q_o^2}\delta(2q_0-2\omega') \propto q_n^*$ always goes to zero at the threshold. Therefore, we focus on the behaviour of the process amplitude. In this context, the arguments of the generalized Bessel functions in the center-of-momentum frame read
\begin{align}
&z_-= -\frac{2\eta}{ms^2}\frac{q\mathrm{sin}(\phi)}{1-\frac{q^2}{\omega'^2} \mathrm{cos}^2(\phi)} \stackrel{s\to s_n}{\to} 0,\\
&z_+ = -\frac{\eta^2}{4s^2}\frac{1}{1-\frac{q^2}{\omega'^2} \mathrm{cos}^2(\phi)} \stackrel{s\to s_n}{\to}  -\frac{\eta^2}{4s_n^2} \equiv z_{+n}
\end{align}
and, consequently, the generalized Bessel function at the threshold can be written as \cite{Reiss, Krainov}
\begin{equation}
\tilde{J}_n(0,z_{+n}) = \begin{cases} J_{n/2}(z_{+n}), & n \textrm{ even,} \\ 0, & n\textrm{ odd. }\end{cases}
\end{equation}
Hence, when an even number of strong field photons $n=2\ell$ is absorbed, the rate (see Eq.~\eqref{ratecmf}) at the threshold behaves
\begin{equation}\label{ratecmfThreshold}
R_{s\to s_n}=\frac{e^2a_0'^2m}{8}\sum_{2\ell\geq n_0}^\infty 
 \frac{J_{\ell}^2\left(-\frac{\eta^2 \ell}{2(1+\eta^2/2)}\right)}{1+\eta^2/2}.
\end{equation}
While this expression tends to vanish as $\eta$ grows ($\eta \gg 1$), the contribution for $\eta \ll 1$ approximates 
\begin{equation}\label{ratecmfThresholdSmallEta}
R^{\eta \ll 1}_{s\to s_n} \approx \frac{e^2a_0'^2m}{8}\sum_{2\ell\geq n_0}^\infty 
\eta^{4\ell}\frac{ \ell^{2(\ell-1)}}{2^{4\ell}\Gamma^2(\ell)},
\end{equation}
where $\Gamma(x)$ stands for the gamma function \cite{NIST}.
We see that the rate above does not vanish. This causes a lifting of the rate at each threshold when an even number of strong field photons is absorbed, which resembles the behaviour at  low intensity of the Breit-Wheeler process in graphene \cite{GolubGraphene}. Our analysis indicates that this peculiarity is a general consequence of the 2+1 dimensionality and does not rely on the Lorentz invariance break down that is caused by the Fermi velocity of this medium.    
\section{Asymptotic study \label{Asymptotics}}
This section is devoted to establishing asymptotic expressions of the Breit-Wheeler pair creation rate (see Eq.~\eqref{rate}) for different intensity parameters of the strong field $\eta$.

\subsection{Behaviour for $\eta \ll 1$ \label{EtaSmall}}
\begin{figure}
\includegraphics[width=8.5cm]{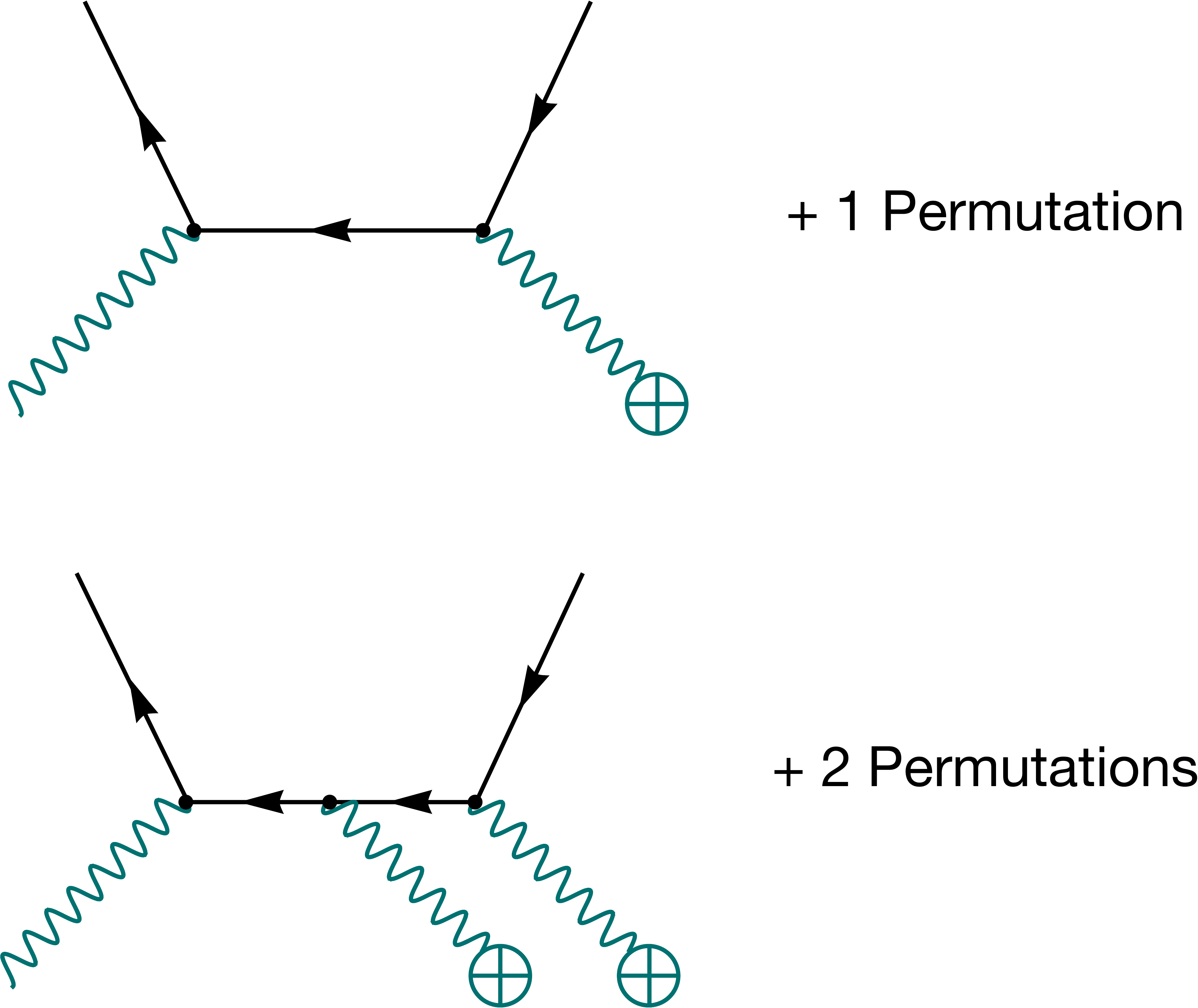}
\caption{\label{FeynPert} Feynman diagrams of two (upper panel) and three (lower panel) photons pair creation processes. In both panels free fermions are represented by external solid lines, internal lines stand for the free fermion propagators, while wavy lines correspond to photons stemming from quantised and classical sources (the latter  are marked by crossed circles). }
\end{figure}
For small laser parameter $\eta \ll 1$ the leading order contribution of Eq.~\eqref{rate} results, when energetically allowed, from the absorption of a single laser photon ($n=1$). A representative Feynman diagram for this process is depicted in the upper panel of Fig.~\eqref{FeynPert}. In this limiting case the effective electron and positron momenta can be simply taken as $q^{\pm}= p^{\pm}$, so that the arguments of the generalized Bessel functions read $z_-=\eta m(\frac{p^-\epsilon}{p^-k}-\frac{p^+\epsilon}{p^+k})$, $z_+=-\frac{\eta^2m^2}{8}\frac{kk'}{kp^+kp^-}$. Their respective proportionality on $\eta$ and $\eta^2$ allows us to expand the Bessel functions (see Appendix C in \cite{Reiss}):
\begin{multline}\label{expansionBessel}
\tilde{J}_n(\eta \beta, \eta^2 \varrho) \approx \left(\frac{\eta^2 \varrho}{2}\right)^{\frac{n}{2}} \\
\quad \times \begin{cases}
\sum_{k=0}^{n/2}\frac{(\beta^2/2\varrho)^k}{(2k)!(n/2-k)!} & \text{for even $n$,} \\
\sum_{k=0}^{(n-1)/2}\frac{(\beta^2/2\varrho)^{k+1/2}}{(2k+1)!((n-1)/2-k)!} & \text{for odd $n$,}
\end{cases}
\end{multline}
where the shorthand notation $\beta=m(\frac{p^-\epsilon}{p^-k}-\frac{p^+\epsilon}{p^+k})$, $\varrho=-\frac{m^2}{8}\frac{kk'}{kp^+kp^-}$ has been introduced. Hence, in the lowest order in $\eta$ the squared amplitude can be written as
\begin{equation}\label{mn1}
|\mathcal{M}^{n=1}_{\eta \ll 1}|^2 = \frac{\eta^2\beta^2}{4}-\frac{\eta^2}{4}\left(1-\frac{(kk')^2}{4kp^+kp^-}\right)
\end{equation}
The use of Eq.~\eqref{gmunu} in combination with the energy-momentum balance $p^++p^-=k+k'$ allows us to express the differential rate of the pair production in the following form
\begin{equation}\label{ratesmalleta}
\begin{split}
&dR^{n=1}_{\eta \ll 1}=\frac{e^2a_0'^2m^2}{8\pi} \delta_{k+k',p^++p^-} |\mathcal{M}^{n=1}_{\eta \ll 1}|^2\frac{d^2p^-d^2p^+}{p_0^-p_0^+},\\
&|\mathcal{M}^{n=1}_{\eta \ll 1}|^2 =\frac{\eta^2}{4}\\
& \qquad \quad \times \left[\frac{(kk')^2}{4kp^+kp^-} -1 +\frac{2m^2kk'}{kp^+kp^-} -\frac{m^4(kk')^2}{(kp^+kp^-)^2}\right].
\end{split}
\end{equation}
It is worth mentioning that this expression coincides with the outcome resulting from perturbation theory when the classical field in Eq.~\eqref{diracEq} is canonically quantised (see details in Appendix~\ref{pairCrePert}).

\begin{figure}
\includegraphics[width=\columnwidth]{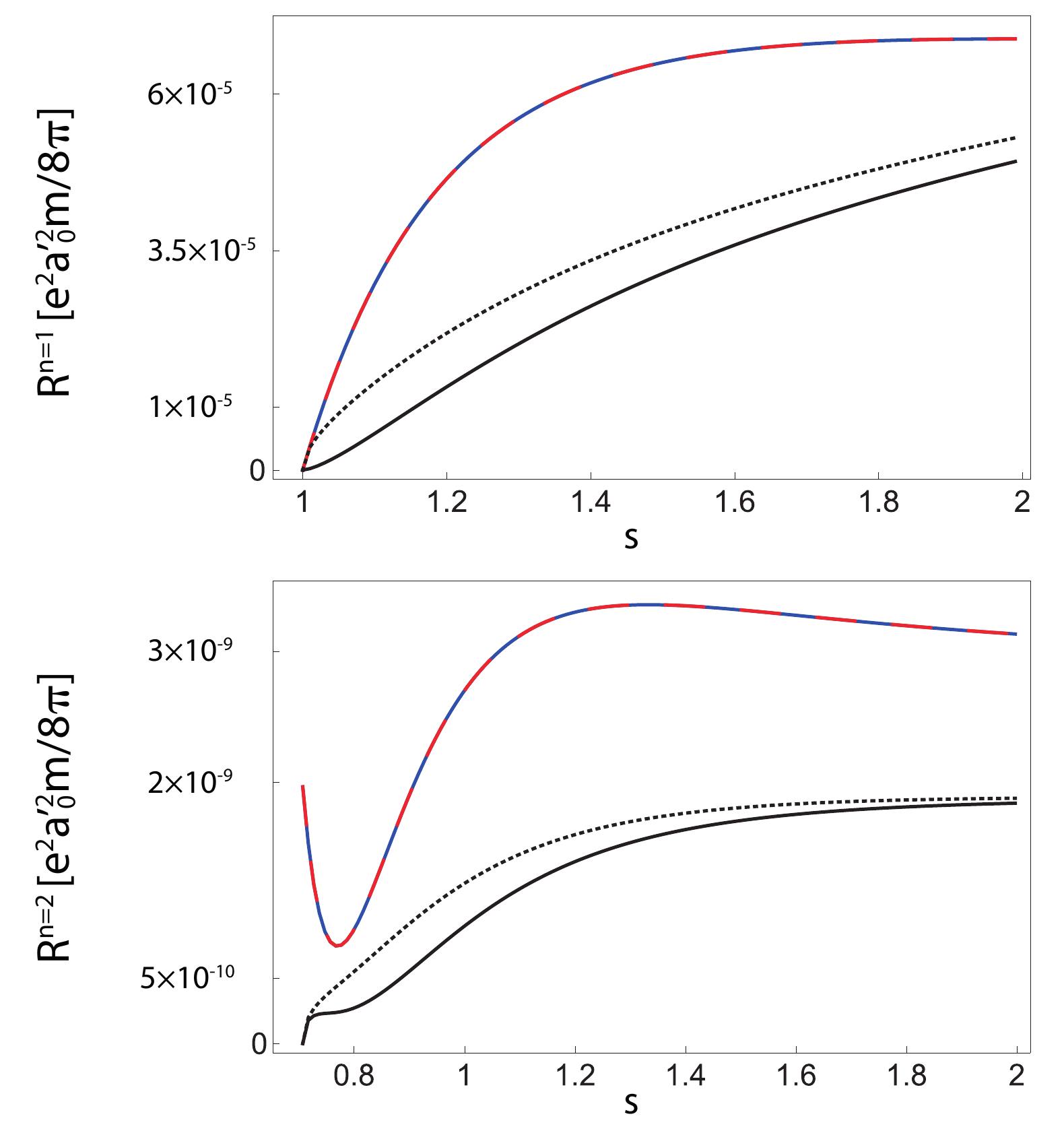}
\caption{\label{fig1} Dependence of the pair production rate in 2+1 dimensions on the Mandelstam invariant $s$ for $\eta=0.01$. In the upper panel the numeric evaluation of the $n=1$ summand of Eq.~\eqref{rate} is depicted in blue and the corresponding asymptotic expression (see \eqref{ratesmalleta1}) is coloured in red. For comparison, the $\mathrm{QED}_{3+1}$ rates for unpolarized and parallel polarized photons \cite{NikishovRitus1} are shown in black solid and black dotted lines correspondingly. 
In the lower panel we compare the outcome of Eq.~\eqref{ratesmalleta2} with the numerically evaluated summand for $n=2$ of Eq.~\eqref{rate} using the same colour scheme as in the upper picture.}
\end{figure}

Finally, we integrate the rate in Eq.~\eqref{ratesmalleta} by adopting polar coordinates. To facilitate the integrations over the momenta we go over to the center-of-momentum frame, where $\vec{p}=\vec{p}^{\ +}=-\vec{p}^{\ -}$, $p_0^+=p_0^-$, $\omega=\omega'$, $kk'=2\omega \omega'=2\omega^2$ and $kp=\omega(p_0-|\vec{p}|\mathrm{cos}(\phi))$. Then, after integrating out $p^-$ and $p_0$ we arrive at
\begin{multline}
R^{n=1}_{\eta \ll 1}=\frac{e^2a_0'^2m^2}{8\pi}\frac{\eta^2}{4} \frac{1}{2\omega}\int_0^{2\pi}d\phi \Big[-1   \\
 -\frac{4u^2}{[1-(1-u)\mathrm{cos}^2(\phi)]^2} + \frac{( 1+4u)}{[1-(1-u)\mathrm{cos}^2(\phi)]} \Big]
\end{multline}  
with $u=m^2/\omega^2$. The remaining integration can be performed analytically by using Eqs.~(3.616.8) and (3.642.3) in Ref.~\cite{Gradshteyn}
\begin{equation}\label{ratesmalleta1}
R^{n=1}_{\eta \ll 1}=\frac{e^2a_0'^2m}{8\pi}\frac{\pi\eta^2}{4s}\left[-1 -\frac{2(s^2+1)}{s^3} + \frac{( 4+s^2)}{s} \right],
\end{equation}
where the result has been expressed in terms of  the Mandelstam variable $s=\sqrt{kk'/2m^2}$. Observe that in this context the energy threshold translates into $s \geq1$. We remark that for $s=1$ the rate above vanishes. In the upper panel of Fig.~\ref{fig1} a comparison between the asymptotic expression in Eq.~\eqref{ratesmalleta1} (red dashed line) and the numerical evaluation for $n=1$ in Eq.~\eqref{rate} (blue curve) is depicted. This assessment has been done by taking $\eta=0.01$. For further comparison, Fig.~\ref{fig1} exhibits the behaviour of the corresponding process in $3+1$ dimensions coloured in black. While the solid line results from an unpolarised photon beam, the dotted one refers to the case of parallel polarisation.

Usually, the derivation of the linear Breit-Wheeler rate is performed within the second order of perturbation theory when considering two quantised photons (see Appendix \ref{pairCrePert}). To bring Eq.~\eqref{ratesmalleta1} into the corresponding form the replacements $a_0^{(\prime)2} \equiv  2/\omega^{(\prime)} A$ need to be done. This way the quantity $\alpha_{2+1}=e^2/(4\pi m)$ appears in the rate, which is the counter part of the fine structure constant in $\mathrm{QED}_{3+1}$. We remark that $\alpha_{2+1}$ plays the role of a perturbative coupling constant provided $ m \gg e^2/(4\pi)$.

Next, we consider the case where two strong field photons are absorbed ($n=2$) (see lower panel of Fig.~\eqref{FeynPert} for the corresponding Feynman diagram). The leading order contribution to this process is described by the amplitude 
\begin{multline}
|\mathcal{M}_{\eta \ll 1}^{n=2}|^2 =\left(\frac{\eta^2 \varrho }{2}+\frac{\eta^2 \beta^2}{8}\right)^2 \\
-\frac{\eta^2}{4}\left(1-\frac{(kk')^2}{4kp^+kp^-}\right)\left(\frac{\eta^2 \beta^2}{8}- \frac{\eta^2 \varrho }{2}\right),
\end{multline} 
where the involved Bessel functions were expanded with help of Eq.~\eqref{expansionBessel}.
Furthermore, we perform the integration in the center-of-momentum frame, which is defined similarly to the case $n=1$ taking into account that $\omega'=2\omega$. Consequently, the  asymptote  of the three-photon reaction reads
\begin{equation}\label{ratesmalleta2}
R^{n=2}_{\eta \ll 1}=\frac{e^2a_0'^2m}{8\pi}\frac{\pi\eta^4}{128s^8}\left[10 - 30 s^2 + 19 s^4 + 12 s^8\right].
\end{equation}
We note that at the threshold $s=s_2=\sqrt{1/2}$ the expression above coincides with the outcome resulting from Eq.~\eqref{ratecmfThresholdSmallEta} for $n=2$ providing a non-zero contribution. This effect can be seen in the lower panel of Fig.~\eqref{fig1}, where the asymptote as given in the equation above is depicted in blue, whereas the red dashed curve results from evaluating Eq.~\eqref{rate} for $n=2$. These outcomes are compared with the $\mathrm{QED}_{3+1}$ rates for unpolarised photon (black solid) and photon polarised parallelly to the classical field (black dotted), which vanish at the threshold.

Finally, we remark that the fact that $\eta$  and $\eta^\prime$ are dimensionless regardless of the spacetime dimension would  allow us to transfer our  predictions for the production rate from the 2+1 dimensional model to a real situation once it is assumed that they coincide with those defined from the physical charge and amplitude in 3+1 dimensions, i.e. if $\eta=\xi$. Observe that, identifying alternatively $\alpha_{2+1}$ with $\alpha\approx1/137$, makes the rate  a function depending explicitly on the field amplitude in 2+1 dimensions. However, the dimension of this quantity $[a_0]=\mathcal{E}^{1/2}$ differs from the one of  its 3+1 dimensional counterpart $\sim \mathcal{E}$, preventing this way a  mutual matching and so, a direct link with physical quantities.

\subsection{Behaviour for $\eta \gg 1 $ \label{EtaLarge}}
Contrary to the previous case of small laser parameters, for $\eta \gg 1$ a large number of summands in Eq~\eqref{rate} will provide non-negligible contributions to the rate.
Indeed, a huge amount of photons is needed for the process to take place so that a transition to the continuum limit $\sum_{n} ... \to \int dn ...$ can be developed.\footnote{Actually, the transition is made on the variable $\mathfrak{s}_n \equiv n/(1+\eta^2/2)=1/s_n^2$ for which the interval between two neighbour terms converges to zero $\Delta \mathfrak{s}_n = (1+\eta^2/2)^{-1} \to d\mathfrak{s}$ as $\eta \to \infty$.}
Afterwards, $n$ together with $\vec{q}^{\ +}$ are integrated out by using the delta functions. Consequently, Eq.~\eqref{rate} reduces to 
\begin{equation}\label{ratelargeeta}
R_{\eta \gg 1}= \frac{e^2a_0'^2m^2}{8\pi}\int \frac{d^2q^-}{q_0^- kq^+} |\mathcal{M}^{n=n^*}_{\eta \gg 1}|^2.
\end{equation}
Here, $|\mathcal{M}^{n=n^*}_{\eta \gg 1}|^2$ is given in Eq.~\eqref{matrixfinal}, where $n \to n^*$ and $n^*=q^-k'/[\omega(k_0'-k_2'-q_0^- + q_2^-)]$. We assume without loss of generality that the strong field $a^\mu$ is polarized in $x_1-$ direction. Next, let us introduce new variables
\begin{equation}
\mathpzc{X} = \frac{kq^+}{m^2}\eta, \ \mathpzc{X}' = \frac{kq^-}{m^2}\eta, \ \gamma^- = q_0^- - q_2^-
\end{equation}
Observe that these definitions allow us to write $z_{\pm}$ (see below Eq.(7)) in the following form
\begin{equation}\label{z}
z_+ =-\frac{\eta^3\kappa}{8\mathpzc{X}\mathpzc{X}'},\
z_- =\frac{\eta^2}{m\mathpzc{X} \mathpzc{X}'}(q^-\epsilon\mathpzc{X} -q^+\epsilon\mathpzc{X}'),
\end{equation}
where the quantum non-linearity parameter in 2+1 dimensions $\kappa = \eta kk'/m^2$ has been introduced.
Further, we perform a substitution $q_2^- \to \gamma^-$ which allows us to write the integral in Eq.~\eqref{ratelargeeta} as
\begin{equation}\label{ratelargeeta1}
R_{\eta \gg 1}= \frac{e^2a_0'^2m^2}{8\pi}\frac{\eta}{m^2}\int_{-\infty}^{\infty} \frac{d q_1^-}{\mathpzc{X}}\int_{0}^{\lambda}\frac{d\gamma^-}{\gamma^-} |\mathcal{M}^{n=n^*}_{\eta \gg 1}|^2,
\end{equation}
where $\lambda=k_0'-k_2'$ and
\begin{equation}\label{largeetaamplitude}
|\mathcal{M}^{n=n^*}_{\eta \gg 1}|^2 = \tilde{J}_{n^*}^2 -\eta^2\left(1-\frac{\kappa^2}{4\mathpzc{X} \mathpzc{X}'}\right)\left[ (\tilde{\mathcal{J}}^1_{n^*})^2 - \tilde{J}_{n^*}\tilde{\mathcal{J}}^2_{n^*}\right].
\end{equation}
The matrix element above still contains generalized
Bessel functions. In order to proceed, we have calculated
their asymptotic expansions in the relevant limits\footnote{In order to exploit the large argument behaviour, the Bessel functions are written in their integral representation $\tilde{J}_n(z_-,z_+) = \frac{1}{\pi}\mathrm{Re}\left[\int_0^{\pi}d\theta  \mathrm{e}^{ f(\theta)}\right]$ with $f(\theta)=-iz_- \mathrm{sin}(\theta)-iz_+ \mathrm{sin}(2\theta) +in\theta$ and the integration is performed asymptotically via the method of steepest descent \cite{SteepestDesc} by expanding $f(\theta)$. Since we are interested in deriving asymptotic formulas for both $\kappa \ll 1$ and $\kappa \gg 1$, an expansion up to the third order term is needed \cite{Leubner}.},
following the lines of Appendix B in Ref.~\cite{NikishovRitus1} and paying
special attention to modifications arising from the reduced
dimensionality. It turns out that the latter induces some
minor changes but does not affect the derivations substantially.
Having this in mind, we find that for $\eta \gg 1$ and $\eta \gg \kappa^{1/3}$ the square of the matrix element approximates
\begin{multline}\label{matrixlargeeta}
|\mathcal{M}^{n=n^*}_{\eta \gg 1}|^2 \approx \frac{2}{\pi}\left(-\frac{1}{4z_+ \mathrm{sin}^2(x_0)}\right)^{2/3} \\
\quad \times \left[\Phi^2(z)-\left(1-\frac{\kappa^2}{4\mathpzc{X} \mathpzc{X}'}\right)\left(\Phi^2(z) + \frac{\Phi'^2(z) }{z}\right) \right]
\end{multline}
with $z=\frac{(-4\sin^2(x_0)z_+)^{2/3}}{\eta^2\sin(x_0)}$. Moreover, $\Phi(z)$ stands for the Airy function \cite{NikishovRitus1}.
Observe that in the equation above a new substitution involving $x_0$ was introduced: $\mathrm{cos}(x_0)=-\frac{\omega}{\kappa m^3}(q_1^-\gamma^+-q_1^+\gamma^-)$ with $\gamma^+ = q_0^+ - q_2^+$, $\gamma^++\gamma^-=\lambda$ and $x_0 \in [0,\pi]$. Furthermore, we substitute $\mathpzc{X}' =\frac{\eta \omega}{m^2}\gamma^- $ and define $\mathpzc{X}'= \frac{\kappa}{2}[1+\mathrm{tanh}(\vartheta)]$, $\mathpzc{X}= \frac{\kappa}{2}[1-\mathrm{tanh}(\vartheta)]$. Notice that the integral over $\mathpzc{X}'$ is defined over the interval $[0, \kappa]$. As the integrand is symmetric in $\mathpzc{X}'$ and $\mathpzc{X}$ with $\mathpzc{X}'+\mathpzc{X} = \kappa$ we can write $\int_0^{\kappa} d \mathpzc{X}'...= 2\int_0^{\kappa/2} d \mathpzc{X}'... $ . Therefore,  the integration over $\vartheta$ will run from $0$ to $\infty$. Having these in mind we obtain
\begin{multline}\label{ratelargeeta2}
R_{\eta \gg 1}\approx \frac{e^2a_0'^2m}{8\pi} \frac{8}{\pi}\int_{0}^{\pi/2} dx_0\int_{0}^{\infty}\frac{d\vartheta}{\mathrm{ch}^2(\vartheta)} \sqrt{z} \\
\times \left[\Phi^2(z)+\mathrm{sh}^2(\vartheta)\left(\Phi^2(z) + \frac{\Phi'^2(z) }{z}\right) \right]
\end{multline}
with $z=\left(\frac{2\mathrm{ch}^2(\vartheta)}{\kappa\mathrm{sin}(x_0)}\right)^{2/3}$.
We remark that using the relations 
$\Phi'^2(z) /z = 1 /(2z) \ d^2\Phi^2(z)/dz^2-\Phi^2(z)$ and $\Phi^2(z) = \frac{1}{2^{2/3}\sqrt{\pi}}\int_0^{\infty}\frac{dt}{\sqrt{t}}\Phi(t+2^{2/3}z) $
combined with the defining equation for the Airy functions \cite{Aspnes} allows us to write the rate as a function depending linearly  on the Airy function of shifted argument $t+2^{2/3}z$
\begin{multline}\label{ratelargeeta3}
R_{\eta \gg 1}\approx \frac{e^2a_0'^2m}{8\pi} \frac{8}{2^{2/3}\pi^{3/2}}\int_{0}^{\pi/2} dx_0 \int_0^{\infty}\frac{dt}{\sqrt{t}}\\
\times\int_{0}^{\infty}\frac{d\vartheta \sqrt{z} }{\mathrm{ch}^2(\vartheta)} 
\left(1+\frac{2^{1/3}\mathrm{sh}^2(\vartheta)}{z}\right) \Phi(t+2^{2/3}z).
\end{multline}

For derivation of asympotes we come back to Eq.~\eqref{ratelargeeta2} and start with considering $\kappa\ll 1$. Since the Airy functions decrease monotonically, the largest contribution to the integral in Eq.~\eqref{ratelargeeta2} results from the region close to $\tilde{\vartheta}=0$ and $\tilde{x}_0=\pi/2$. Hence, we expand $\Phi$ and $\Phi'$ for large arguments  as $1/\kappa \gg 1$ 
\begin{align}
&\Phi^2(z) =\frac{z}{3\pi} K^2_{1/3}\left(\frac{2}{3}z^{3/2}\right)\overset{z\to \infty}{\longrightarrow} \frac{z^{-1/2}}{4\pi}\mathrm{e}^{-\frac{4}{3}z^{3/2}},\\
&\Phi'^2(z) = \frac{z^2}{3\pi} K^2_{2/3}\left(\frac{2}{3}z^{3/2}\right)\overset{z\to \infty}{\longrightarrow}\frac{z^{1/2}}{4\pi}\mathrm{e}^{-\frac{4}{3}z^{3/2}}
\end{align}
with the modified Bessel functions of the second kind $K_\nu(x)$ \cite{NIST}.
Taking into account the expansions above we further expand all involved functions in $\tilde{\vartheta}$ and $\tilde{x}_0$ and perform the remaining integrations, ending up with
\begin{equation}\label{ratesmallkappa}
R_{\eta \gg 1, \kappa\ll 1}\approx \frac{e^2a_0'^2m}{8\pi}\frac{3\kappa}{8\sqrt{2}}\mathrm{e}^{-\frac{8}{3\kappa}}.
\end{equation}
Comparing the equation above with its 3+1 dimensional analogue (as given in Eq.~(36') of Ref.~\cite{NikishovRitus1}) we see that the power linked to the quantum non-linearity parameter $\kappa$ in the pre-exponential factor is equal to $1$ rather than $3/2$. This difference comes from the reduced dimensionality of the phase space: Similarly as in $\mathrm{QED}_{3+1}$, the integrand in this limit depends on $\kappa$ solely through the exponential function $\mathrm{e}^{-\frac{4}{3}z^{3/2}}$ and integration over each variable provides a factor $\kappa^{1/2}$. Since in 2+1 dimensions $z$ does not depend on
a variable responsible for a dimension orthogonal to the plane spanned by photon polarisation and propagation vectors, we obtain a power of $1/2$ less in the quantum non-linearity parameter than in 3+1 dimensions.

Establishment of the asymptotic formula for $\kappa \gg 1$ demands a more elaborated procedure. To this end, we substitute $\kappa \mathrm{sin}(x_0) = p$ and $\mathrm{ch}(\vartheta) = u$ in Eq.~\eqref{ratelargeeta2}:
\begin{multline}
R_{\eta \gg 1, \kappa\gg 1}\approx \frac{e^2a_0'^2m}{8\pi}\frac{16}{3\pi^2}\int_0^\kappa\frac{dp}{p\sqrt{\kappa^2-p^2}} \int_1^\infty \frac{du}{\sqrt{u^2-1}} \\
\times \left[ u^2 K_{1/3}^2\left(\frac{4u^2}{3p}\right) + (u^2-1)K_{2/3}^2\left(\frac{4u^2}{3p}\right)\right]
\end{multline}  
and split the integration over $p$ into two intervals: $[0,p_0]$ and $[p_0, \kappa]$ with $p_0 \ll \kappa$. 
Then, the contribution from the first interval will be negligible as it scales with $\kappa^{-1}$: 
\begin{equation}
\int_0^{p_0}\frac{dp}{p\sqrt{\kappa^2-p^2}} ... \approx \frac{1}{\kappa}\int_0^{p_0}\frac{dp}{p} ... \sim \frac{p_0^{4/3} }{\kappa} \ll 1.
\end{equation}
For the extant integral we perform a transition to a new variable $t=4u^2/3p$, whereas $p$ remains unchanged. As a consequence, 
\begin{multline}\label{Retag1kappag1}
R_{\eta \gg 1, \kappa\gg 1}\approx \frac{e^2a_0'^2m}{8\pi}\frac{4}{\sqrt{3}\pi^2}\int_{p_0}^\kappa\frac{dp}{\sqrt{p}\sqrt{\kappa^2-p^2}} 
\int_{\frac{4}{3p}}^\infty dt \\
\times \left[ \frac{3p}{4}\sqrt{\frac{t}{\frac{3pt}{4}-1}}K_{1/3}^2\left(t\right) + \sqrt{\frac{\frac{3pt}{4}-1}{t}}K_{2/3}^2\left(t\right)\right].
\end{multline}

At this point it is suited to split the integration over $t$ into two sectors: 
\begin{equation}\label{intInt}
\int_{4/3p}^{\infty} dt ... = \int_{4/3p}^{t_0} dt ... + \int_{t_0}^{\infty} dt ... \ .
\end{equation}
Here, the parameter $t_0$ satisfies the conditions
$4/3\kappa \ll 4/3p_0 \ll t_0 \ll 1, \quad 3\kappa t_0/4 \gg 3p_0 t_0/4 \gg1$.
Let us consider firstly the contribution defined over $[t_0,\infty]$. In this region $3pt/4 \gg 1$ holds  and we obtain approximately
\begin{multline}\label{ratelargeetaR1}
 \int_{t_0}^{\infty} dt ...\approx\int_{t_0}^\infty dt \sqrt{\frac{3p}{4}}
\left( K_{1/3}^2\left(t\right) + K_{2/3}^2\left(t\right)\right) \\
\approx \sqrt{\frac{3p}{4}} \frac{3 \Gamma^2(\frac{2}{3})}{2^{2/3}t_0^{1/3}}
\end{multline}
where the smallness of $t_0$ has been used [$t_0\ll1$] and $\Gamma(x)$ denotes the gamma function \cite{NIST}. 

Now, let us turn our attention to the contribution in Eq.~\eqref{intInt} defined over $[4/3p,t_0]$. As in this interval $t \ll 1$, we can use the small argument behaviour  of the modified Bessel functions $K_{\nu}(t) \approx \Gamma(\nu)/2(t/2)^{\nu}$ \cite{NIST} and, consequently, perform the integration over $t$. Taking advantage of the fact that $4/3pt_0 \ll 1$ and $t_0 \ll 1$ we arrive at
 \begin{multline}\label{ratelargeetaR2}
\int_{4/3p}^{t_0} dt ... \approx 
 - \sqrt{\frac{3p}{4}}\frac{3 \Gamma^2(\frac{2}{3})}{2^{2/3}t_0^{1/3}} + \frac{3^{5/6} \sqrt{\pi} }{2^{10/3}}\frac{\Gamma^2(\frac{2}{3}) \Gamma(\frac{1}{3})}{\Gamma(\frac{11}{6})}p^{5/6}  \\
 +\frac{3^{1/6} \sqrt{\pi} }{ 2^{5/3}}\frac{\Gamma^2(\frac{1}{3}) \Gamma(-\frac{1}{3})}{\Gamma(\frac{1}{6})}p^{1/6} .
\end{multline}
We combine both Eqs.~\eqref{ratelargeetaR1} and \eqref{ratelargeetaR2} in Eq.~\eqref{intInt}. As a consequence, $ t_0$ drops out as it should and the resulting expression is inserted into Eq.~\eqref{Retag1kappag1}. Hence, to the leading order in $\kappa$ we obtain 
\begin{equation}\label{ratelargekappa1}
R_{\eta \gg 1, \kappa\gg 1}\approx \frac{e^2a_0'^2m}{8\pi}\frac{3^{11/6}}{5\ 2^{1/3} \pi } \Gamma^2\left(\frac{2}{3}\right) \kappa^{1/3}.
\end{equation}
As for the case $\kappa \ll 1$, a deviation with respect to $\mathrm{QED}_{3+1}$ occurs.
Indeed, the presence of the $\kappa^{1/3}$ dependence  is in sharp contrast to the well established $\chi^{2/3}$ behavior that the rate in 3+1 dimensions  manifests (for comparison see Eq.~(36'') in Ref.~\cite{NikishovRitus1}). It  provides the first evidence that  the proper expansion parameter for $\mathrm{QED}_{2+1}$ in  the strong  field  (see Eq. (1))  could be   $g\sim \alpha_{2+1}\kappa^{1/3}$, whenever $g\ll1$.  However,  at this point,  this statement must be considered as a conjecture;  its confirmation will require  a more elaborated   investigation of consecutive terms within the perturbative expansion  \cite{Mironov, Podszus,Ilderton,NikishovRitus3}. While this analysis is beyond the scope of the present study, some interesting consequences can be anticipated owing to the superrenormalisable feature of the theory.  Firstly, upon identifying $\kappa\sim \chi\gg1$,  the presence of the  $1/3$ exponent  would induce a breakdown of perturbation theory softer  than in strong-field $\mathrm{QED}_{3+1}$ if  $\alpha_{2+1}\sim \alpha$. However, if $\alpha_{2+1}\gg \alpha \kappa^{1/3}$, i.e.  for a mass scale of the electric  charge   $e^2/(4\pi)\gg m \alpha \kappa^{1/3} $, this break down in $\mathrm{QED}_{2+1}$ could  be stronger than in $\mathrm{QED}_{3+1}$.

\begin{figure}
\includegraphics[width=\columnwidth]{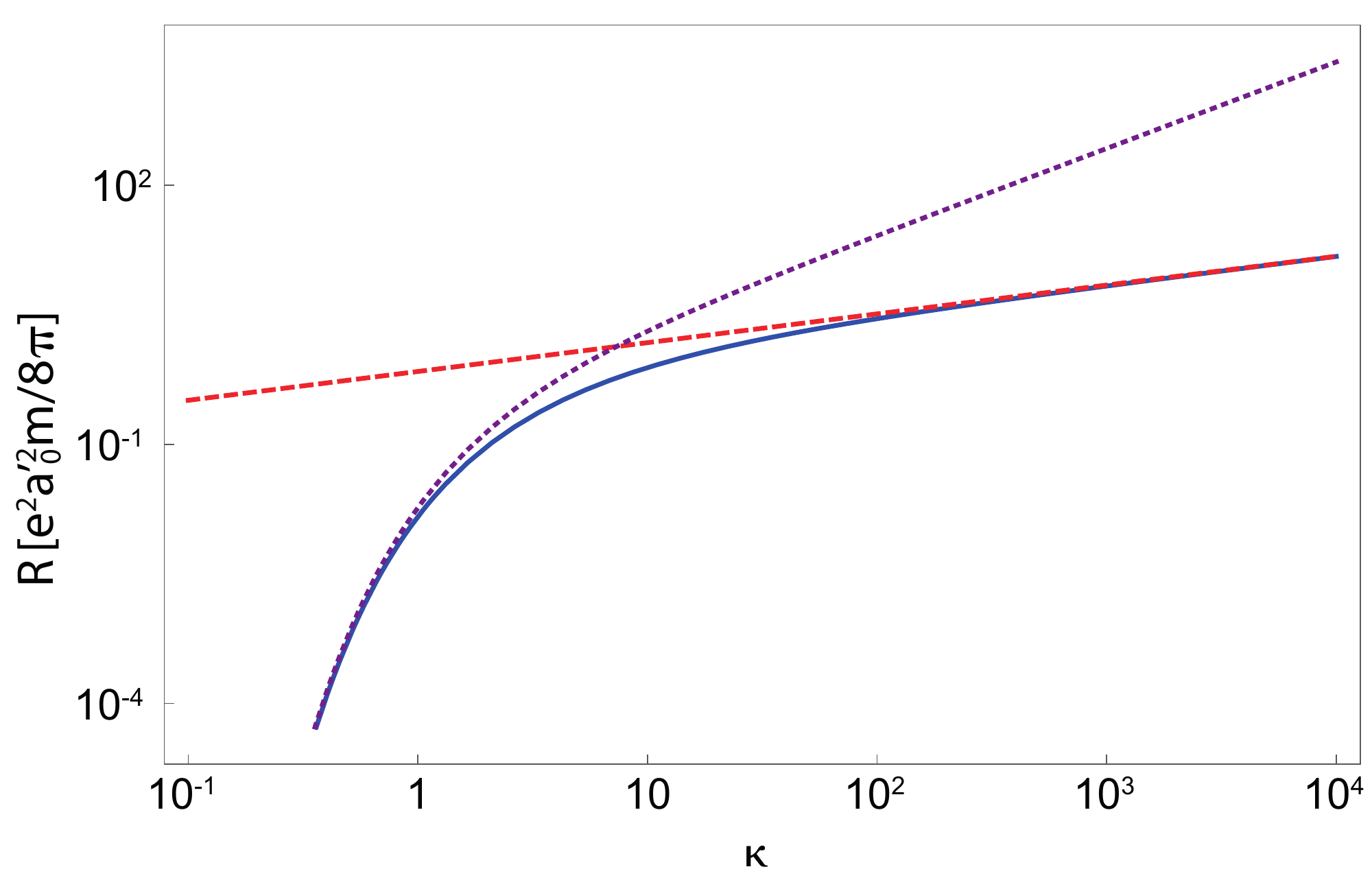}
\caption{\label{fig2} Dependence of the Breit-Wheeler pair production rate in $\mathrm{QED}_{2+1}$ on the quantum non-linearity parameter $\kappa$ [blue solid line] at asymptotically large value of $\eta$. The corresponding behaviour of $R$ for $\kappa \ll 1$ (see Eq.~\eqref{ratesmallkappa}) and $\kappa \gg 1$  (see Eq.~\eqref{ratelargekappa1})  are plotted in purple dotted and red dashed lines, respectively.} 
\end{figure}

The result of this section is summarised in Fig.~\ref{fig2}, which shows the dependence of $R$ on the quantum non-linearity parameter $\kappa$ [blue solid line]. This figure depicts in addition the asymptotes for small $\kappa$ [purple dotted line] and large $\kappa$ [red dashed line] given in Eqs.~\eqref{ratesmallkappa} and \eqref{ratelargekappa1} correspondingly.

\section{Conclusion}\label{conclus}

In the present work we have investigated the Breit-Wheeler pair creation in 2+1 dimensional Minkowski space. Special attention was paid to the domains of small and large intensity parameter $\eta$. We have shown that the dimensionality plays a crucial role when considering the process rates leading to surprising differences when comparing them to the 3+1 dimensional case. We have found that for $\eta\ll 1$ the low dimensionality causes non-vanishing contributions to the rate at the threshold when an even number of strong field photons is absorbed. This outcome supports and extends the results obtained in a recent study on bandgaped graphene \cite{GolubGraphene}. As this feature persists in a Lorentz invariant vacuum, we conclude that in a honeycomb of carbon atoms it occurs independently of the medium.

In the high intensity regime ($\eta \gg 1$), the 2+1 dimensionality induces changes in the pair production rate which are manifested through a dependence on the quantum non-linearity parameter $\kappa$ different from the one appearing within the pure $\mathrm{QED}_{3+1}$ context. On the basis of this finding we have argued that the so far accepted criteria within the strong field $\mathrm{QED}_{3+1}$ community on the applicability of perturbation theory might be subject to modifications as the spacetime dimensionality changes. Indeed, we have seen that the super-renormalisable  character of $\mathrm{QED}_{2+1}$ constitutes an aspect to take into account  in this regard  because, once the electron mass becomes smaller than the energy scale linked to the square of the electric charge, the effective fine-structure constant  is no longer a small parameter and a break down of perturbation theory could take place even for quantum non-linearity parameter of the order of unity. 

\begin{acknowledgments}
This work has been funded by the Deutsche Forschungsgemeinschaft (DFG) under Grant No. 416699545 within the Research Unit FOR 2783/1 and Grant No. 388720772 (MU 3149/5-1).
\end{acknowledgments}

\appendix

\section{Pair creation by two photons in 2+1 dimensions \label{pairCrePert}}
In 2+1 dimensions the S-matrix element which describes the electron positron pair creation out of two photons looks similarly to its counterpart in $\mathrm{QED}_{3+1}$ 
\begin{equation}
S_{fi}=-\frac{ie^2}{A^2}\sqrt{\frac{m^2}{p_0^+p_0^-}} \sqrt{\frac{1}{2^2\omega\omega'}}(2\pi)^3\delta_{k+k',p^++p^-}\mathcal{M}_{p}
\end{equation}
with 
\begin{equation}\label{matrixpert}
\mathcal{M}_{p}=\bar{u}_{p^-}\left( \frac{\slashed{\epsilon}\slashed{\epsilon}'}{\slashed{k}'-\slashed{p}^+-m+i0} + \frac{\slashed{\epsilon}'\slashed{\epsilon}}{\slashed{k}-\slashed{p}^+-m+i0}\right)v_{p^+} .
\end{equation}
Here, the fermion propagator in 2+1 dimensions reads \[S_F(x-y)=\int \frac{d^3 p}{(2\pi)^3}\frac{\mathrm{e}^{ip(x-y)}}{\slashed{p}-m+i0}.\]

The probability rate associated with the perturbative pair creation results from Eq.~\eqref{rateinitial} when substituting $q^- \to p^-$ and $q^+ \to p^+$. Explicitly, 
\begin{equation}\label{rateP}
dR_{\gamma\gamma' }=\frac{e^4}{A^2}\frac{m^2}{8\pi \omega\omega'}\delta_{k+k',p^++p^-}|\mathcal{M}_{p}|^2\frac{d^2p^-}{p^-_0}\frac{d^2p^+}{p^+_0},
\end{equation}
where the square of the S-matrix element is
\begin{multline}
|\mathcal{M}_{p}|^2 = \frac{1}{16m^2}\mathrm{Tr}\Big[ \\
 \left(\slashed{\epsilon}\frac{\slashed{k}'-\slashed{p}^++m}{k'p^+}\slashed{\epsilon}' +\slashed{\epsilon}'\frac{\slashed{k}-\slashed{p}^++m}{kp^+}\slashed{\epsilon} \right)  (\slashed{p}^+ -m) \\
\times \left(\slashed{\epsilon}'\frac{\slashed{k}'-\slashed{p}^++m}{k'p^+}\slashed{\epsilon} + \slashed{\epsilon}\frac{\slashed{k}-\slashed{p}^++m}{kp^+}\slashed{\epsilon}'\right) (\slashed{p}^- +m)\Big].
\end{multline}
Using the properties of the gamma matrices (see footnote \ref{footnote4}) and the energy-momentum conservation, the expression above reduces to
\begin{equation}\label{matrixperp}
|\mathcal{M}_{p}|^2 = \frac{1}{m^2}\left[\frac{(kk')^2}{4kp^+kp^-} -1 +\frac{2m^2kk'}{kp^+kp^-} -\frac{m^4(kk')^2}{(kp^+kp^-)^2}\right].
\end{equation}
We note that the form of $|\mathcal{M}_{p}|^2$ differs from its 3+1 dimensional counterpart (see \cite{Greiner}). When inserting Eq.~\eqref{matrixperp} into Eq.~\eqref{rateP} we obtain the differential rate
\begin{multline}
dR_{\gamma\gamma' }=\frac{e^4a_0^2a_0'^2}{32\pi}\delta_{k+k',p^++p^-}\frac{d^2p^-}{p^-_0}\frac{d^2p^+}{p^+_0} \\
\times \left[\frac{(kk')^2}{4kp^+kp^-} -1 +\frac{2m^2kk'}{kp^+kp^-} -\frac{m^4(kk')^2}{(kp^+kp^-)^2}\right],
\end{multline}
where the relation $a_0^{(\prime)2} \equiv  2/\omega^{(\prime)} A$ has been used. This expression coincides with Eq.~\eqref{ratesmalleta} provided the intensity parameter $\eta=e a_0/m$ is identified.

\end{document}